\documentclass[11pt,prd,aps,nofootinbib]{revtex4}

\usepackage{epsf,graphicx}
\usepackage{times,latexsym,euscript,amstext,amssymb,amsbsy,amsmath,amsfonts}
\usepackage{epsfig}
\usepackage[T1]{fontenc}
\usepackage{charter}
\usepackage{wrapfig}
\usepackage[a4paper,portrait,top=1.5cm, bottom=2.2cm,left=1.5cm,right=1.5cm]{geometry}
\usepackage[colorlinks,citecolor=blue,linktoc=all,linkcolor=cyan]{hyperref}
\newcommand{\be}{\begin{equation}}
\newcommand{\ee}{\end{equation}}
\newcommand{\bea}{\begin{eqnarray}}
\newcommand{\eea}{\end{eqnarray}}

\def\barr{\left(\begin{array}{c}}
\def\earr{\end{array}\right)}
\def\bmat{\left(\begin{array}{cc}}
\def\emat{\end{array}\right)}
\begin{document}

\title{Accessing the real part of the forward $J/\psi - p$ scattering amplitude from $J/\psi$~photo-production on protons around threshold}

\author{Oleksii Gryniuk}
\author{Marc Vanderhaeghen}
\affiliation{Institut f\"ur Kernphysik \& PRISMA  Cluster of Excellence, Johannes Gutenberg Universit\"at,  D-55099 Mainz, Germany}

\date{\today}

\begin{abstract}

We provide an updated analysis of the forward $J/\psi$-p scattering amplitude, 
relating its imaginary part to 
$\gamma p \to J/\psi p$ and $\gamma p \to c \bar c X$ cross section data, and calculating its real part through a 
once-subtracted dispersion relation. From a global fit to both differential and total cross section data, we extract a value for the 
spin-averaged $J/\psi$-p s-wave scattering length $a_{\psi p} = 0.046 \pm 0.005$~fm, which can be translated into a 
$J/\psi$ binding energy in nuclear matter of $B_\psi = 2.7 \pm 0.3$~MeV. 
We estimate the forward-backward asymmetry to the $\gamma p \to e^- e^+ p$ process around the $J/\psi$ resonance, which results from interchanging the leptons in the interference between the $J/\psi$ production and the Bethe-Heitler mechanisms. 
We show that to good approximation this asymmetry depends linearly on $a_{\psi p}$, and can reach values around -25\% 
for forthcoming $J/\psi$ threshold production experiments at Jefferson Lab.  Its measurement can thus provide a very sensitive observable for a refined extraction of $a_{\psi p}$.  
  
\end{abstract}

\maketitle
\thispagestyle{empty}
%\tableofcontents

\section{Introduction}

The interaction between heavy quarkonia, such as $J/\psi$, and protons or nuclear matter has generated a lot of interest over the past decades as it provides a unique setting to test the gluonic van der Waals interaction in Quantum Chromo Dynamics (QCD). As valence light-quark exchange between the $J/\psi$ and the nuclear system cannot occur, and light quark anti-quark exchange is strongly suppressed due to the Okubo-Zweig-Iizuka (OZI) rule, the interaction proceeds dominantly through multiple gluon exchange which provides a weakly attractive interaction. 
As the heavy quarkonium is a small size system, it can be treated as a color dipole, and its interaction with the nucleon or nucleus may be estimated from the knowledge of its chromo-electric polarizability, see 
Refs.~\cite{Kharzeev:1995ij,Voloshin:2007dx,Hosaka:2016ypm} for reviews and references therein. 
For nuclei, this attraction may be strong enough to provide a bound state between the
$c \bar c$ state and the nucleus \cite{Brodsky:1989jd, Wasson:1991fb, Luke:1992tm}. 
Using a perturbative calculation for the chromo-electric polarizability of a heavy quarkonium~\cite{Peskin:1979va} 
and a two-gluon exchange interaction, 
Ref.~\cite{Luke:1992tm} estimated a binding energy for a $J/\psi$ in nuclear matter $B_\psi \sim 10$~MeV.  
In subsequent works, also higher-order non-perturbative modifications to the interaction between quarkonia and nucleons and nuclei were explored, including the coupling of the $J/\psi$ to $D \bar D$, $D \bar D^\ast$ and $D^\ast \bar D^\ast$ intermediate 
states~\cite{Brodsky:1997gh,Ko:2000jx,Tsushima:2011kh}, which may also lead to attractive interactions. The latter work~\cite{Tsushima:2011kh} predicted $J/\psi$ binding energies ranging from around 5 MeV in $^4$He to around 15 - 20 MeV in $^{208}$Pb. 

The quantitative study of a possible formation of such bound states requires a more precise knowledge of the $J/\psi$-nucleon interaction at low energies, which may be characterized by its (spin-independent) s-wave scattering length 
$a_{\psi p}$, 
corresponding to a $J/\psi$-proton (p) total cross section at threshold of $\sigma_{\psi p} \equiv 4 \pi a_{\psi p}^2$.  
In the absence of a $J/\psi$-p bound state, a small positive (negative) value of $a_{\psi p}$ 
would indicate a weakly attractive 
(repulsive) $J/\psi$-p interaction. If the attraction is sufficiently strong, it may then support $J/\psi$-nuclear bound states~\cite{Yokota:2013sfa}.
It has been estimated using QCD sum rules~\cite{Hayashigaki:1998ey} that $a_{\psi p} \sim 0.1$~fm, corresponding with 
$\sigma_{\psi p} \sim 1.26$~mb. Calculations based on the rather uncertain value of the $J/\psi$ chromo-electric 
polarizability~\cite{Lakhina:2003pj} provide estimates for $a_{\psi p}$ ranging between a value of $a_{\psi p} = 0.05$~fm~\cite{Kaidalov:1992hd} at the lower end, and a value of 
$a_{\psi p} = 0.37$~fm~\cite{Sibirtsev:2005ex}  at the higher end. The latter value would lead to $J/\psi$ binding energy in nuclear matter exceeding 20 MeV. 

In recent years, the question whether $J/\psi$-nuclear bound states exist 
became also amenable to lattice QCD calculations~\cite{Yokokawa:2006td,Kawanai:2010ev,Beane:2014sda}. The most recent of these studies~\cite{Beane:2014sda}, inferred a charmonium-nuclear matter binding energy $B_\psi \lesssim 40$~MeV. The current  
lattice calculations were however performed at large pion masses ($m_\pi \sim 805$~MeV), and it is possible that the systems involving the lightest nuclei will therefore be unbound at the physical pion mass.  Future calculations for smaller quark masses are clearly called for. 
   
The $J/\psi$-p interaction was furthermore studied in Refs.~\cite{Kharzeev:1994pz,Kharzeev:1998bz} 
as a probe of the color deconfinement in high-energy nucleus-nucleus collisions. Based on the small size of the $J/\psi$, around $r_\psi \sim 0.2$~fm $\ll \Lambda^{-1}_{QCD}$, the $J/\psi$-p cross section was estimated in those works in terms of the gluon distribution in the nucleon and related through Vector Meson Dominance (VMD) to the $J/\psi$ 
photo-production cross section on the proton.  
Using different parameterizations of the gluon distributions in a proton from deep inelastic scattering, the experimental behavior of the cross section was well reproduced in Ref.~\cite{Kharzeev:1998bz}. 
In a related work~\cite{Redlich:2000cb},  by using VMD and using data for hidden and open charm photo-production on a proton as input, a  phenomenological estimate of the $J/\psi$-p elastic cross section was given. 
In the present work, we will provide an update along these lines, with the aim to provide 
an improved extraction of the threshold $J/\psi$-p scattering amplitude. For this purpose we will evaluate the $J/\psi$-p forward scattering amplitude in a dispersive formalism. As in Ref.~\cite{Redlich:2000cb}, we will constrain the imaginary part from the present world data of hidden and open-charm photo-production.  The real part will be evaluated through a dispersion relation, which involves one subtraction constant, which can be related to $a_{\psi p}$. 
We will be able to substantially improve the precision of the fit, 
by including the forward differential cross section data for $\gamma p \to J/\psi p$ in the fit, 
and study the sensitivity of the $\gamma p \to J/\psi p$ cross section in the threshold region to the subtraction constant.  
This will then allow us to make a quantitative study for planned experiments  
at the Jefferson Laboratory (JLab)  \cite{SoLID, HallB, HallC}, which are aimed to measure the $\gamma p \to J/\psi p$ process in the threshold region.   

The paper is organized as follows.  We describe the forward $J/\psi$-p scattering amplitude in Section 2, relating  
its imaginary part to $\gamma p \to J/\psi p$ and $\gamma p \to c \bar c X$ data. We subsequently calculate the real part from a dispersion relation, involving one subtraction constant. In our framework, the 6 parameters describing the discontinuities, and the one subtraction constant are obtained from a global fit to both total and forward differential 
photo-production cross sections.  
In Section 3, we then start from this $\gamma p \to J/\psi p$ amplitude to describe the $\gamma p \to J/\psi p \to e^- e^+ p$ process, and calculate the forward-backward asymmetry for the $\gamma p \to e^- e^+ p$ process around the $J/\psi$ resonance, which results from interchanging the leptons in the interference between the $J/\psi$ production mechanism and the competing Bethe-Heitler mechanism. We will show that this forward-backward asymmetry, which is 
proportional to the real part of the $J/\psi$-p amplitude, provides a very sensitive observable to extract the subtraction constant in the forward $J/\psi$-p scattering amplitude, which in turn allows to extract $a_{\psi p}$. We will present results for this forward-backward asymmetry, including error bands resulting from our fitting procedure, in the kinematics of planned experiments at JLab. Finally we will provide our conclusions in Section 4.

\section{Forward $J/\psi - p$ scattering amplitude and $\gamma \, p \to J/\psi \, p$ process}

We consider the forward $J/\psi$-p  elastic scattering process, 
which is described by the spin-averaged forward scattering amplitude $T_{\psi p}(\nu)$, where the shorthand $\psi$ denotes the $J/\psi$ state. 
The amplitude $T_{\psi p}$ depends on the crossing variable $\nu$, defined in terms of the Mandelstam invariants as:
\bea
\nu \equiv \frac{s - u}{4} = \frac{1}{2} (s - M^2 - M_\psi^2),
\eea
where $M (M_\psi)$ stand for the masses of the proton $(\psi)$ respectively. 
  
The forward differential cross section for the $\psi \, p \to \psi \, p$ process can then be expressed as:
\bea
\frac{d \sigma}{dt} \biggr|_{t = 0} (\psi p \to \psi p) = \frac{1}{64 \, \pi \, s \, q_{\psi p}^2} \, \big| T_{\psi p}(\nu) \big|^2,
\eea
where in the forward direction the momentum transfer $t = 0$, and where $q_{\psi p}$ denotes the magnitude of the $\psi$ three-momentum in the c.m. frame, given by:
 \bea 
 q_{\psi p}^2  = \frac{1}{4 s} \left[ s - (M_\psi + M)^2 \right] \left[ s - (M_\psi - M)^2 \right].
 \eea 
  
The optical theorem relates the imaginary part of  $T_{\psi p}(\nu)$ to the $\psi \, p \to X$ total cross section $\sigma_{\psi p}^{tot}$ as:
\bea
\Im T_{\psi p}(\nu) = 2 \sqrt{s} \, q_{\psi p} \, \sigma_{\psi p}^{tot}(\nu).
\label{eq:opt}
\eea
The amplitude $T_{\psi p}(\nu)$ has the property that it is even under crossing, i.e. $T_{\psi p}(-\nu) = T_{\psi p}(\nu)$.
The real part of the amplitude $T_{\psi p}(\nu)$ can be reconstructed from the knowledge of the imaginary part along the real $\nu$-axis using a dispersion relation, provided the integral converges. For large $\nu$, the amplitude is diffractive following approximately the behavior $\Im T_{\psi p}(\nu) \sim \nu^a$, with $1 \leq a < 2$. The convergence of the dispersion integral therefore requires one subtraction. This leads to the subtracted dispersion relation:
\bea
\Re T_{\psi p}(\nu) = T_{\psi p}(0) + \frac{2}{\pi} \nu^2 \int_{\nu_{el}}^\infty d \nu^\prime \frac{1}{\nu^\prime} \frac{\Im T_{\psi p}(\nu^\prime)}{\nu^{\prime \, 2} - \nu^2},
\label{eq:disp}
\eea
where $\nu_{el} \equiv M M_\psi$, corresponds with the elastic threshold $s = s_{el} = (M_\psi + M)^2 = 16.28$~GeV$^2$. 
Furthermore in Eq.~(\ref{eq:disp}), the real subtraction constant $T_{\psi p}(0)$ denotes the amplitude at $\nu = 0$. 
This subtraction constant can be predicted in models, see e.g.~\cite{Kharzeev:1998bz}, or has to be obtained from lattice QCD. 
Alternatively, we can use it as a fit parameter and extract it from data. In the following, we will show the sensitivity to extract 
this subtraction constant from the measurement of the $\gamma \, p \to \psi \, p$ process in the threshold region, 
and will relate it with the $\psi$-p s-wave scattering length $a_{\psi p}$.

Physically, the discontinuity of the amplitude $T_{\psi p}(\nu)$ entering the integrand of Eq.~(\ref{eq:disp}) has two contributions: 
an elastic cut starting at $s_{el}$, and an inelastic contribution  corresponding with 
open charm (meson) production on the proton. We will parameterize the inelastic contribution to $T_{\psi p}$ by an effective inelastic cut  which starts at the $D \bar D$ meson production threshold, corresponding with $s_{inel} = (M + 2 M_D)^2 = 21.79$~GeV$^2$, or equivalently $\nu_{inel} = 5.66$~GeV$^2$. The 
imaginary part of $T_{\psi p}$ is then obtained as sum of elastic and inelastic discontinuities:
\bea
\Im T_{\psi p}(\nu)  = \theta(\nu - \nu_{el}) \,  {\rm Disc}_{el} T_{\psi p}(\nu) +   \theta(\nu - \nu_{inel}) \,  {\rm Disc}_{inel} T_{\psi p}(\nu).
\label{eq:disctot}
\eea

We will parameterize the elastic and inelastic discontinuities by the following 3-parameter forms:
\bea
{\rm Disc}_{el} T_{\psi p}(\nu)  &=& 
C_{el} \left( 1 - \frac{\nu_{el}}{\nu} \right)^{b_{el}}  \left( \frac{\nu}{\nu_{el}} \right)^{a_{el}} 
\label{eq:discel1} \\
{\rm Disc}_{inel} T_{\psi p}(\nu)  &=& 
C_{inel} \left( 1 - \frac{\nu_{inel}}{\nu} \right)^{b_{inel}}  \left( \frac{\nu}{\nu_{inel}} \right)^{a_{inel}}, 
\label{eq:discinel1}  
\eea
where the factors $\sim (1 - \nu_{thr} / \nu)^b$  
determine the behavior around the respective threshold $\nu_{thr}$, and the 
factors  $\sim \nu^a$ determine the Regge behavior of the amplitude at large $\nu$. 
In the following we will discuss how we can determine the respective parameters 
appearing in the elastic and inelastic discontinuities. 

The discontinuity across the elastic cut, ${\rm Disc}_{el}$, 
is related through the optical theorem to the $\psi \, p \to \psi \, p$ elastic scattering cross section 
$\sigma_{\psi p}^{el}$ as~:
\bea
{\rm Disc}_{el} T_{\psi p}(\nu) =  2 \sqrt{s} \, q_{\psi p} \, \sigma_{\psi p}^{el}.
\label{eq:discel2}
\eea
We will use the vector meson dominance (VMD) assumption to relate the elastic cross section 
$\sigma_{\psi p}^{el}$ to the $\gamma p \to \psi p$ cross section~\cite{Barger:1975ng,Redlich:2000cb}: 
\bea
\sigma_{\psi p}^{el} = \left( \frac{M_\psi}{e f_\psi} \right)^2 \left( \frac{q_{\gamma p}}{q_{\psi p}} \right)^2 \, \sigma (\gamma p \to \psi p), 
\label{eq:sigmael}
\eea
with electric charge $e$ given through $\alpha = e^2 / (4 \pi) \simeq 1/137$, and  where $f_\psi$ is the $\psi$ decay constant, which is obtained from the $\psi \to e^+ e^-$ decay as 
\begin{eqnarray}
\Gamma_{\psi \to ee} = \frac{4 \pi \alpha^2}{3} \frac{f_\psi^2}{M_\psi}.
\end{eqnarray}
The experimental value $\Gamma_{\psi \to ee} =  5.55$~keV yields $f_\psi = 0.278$~GeV. Furthermore, $q_{\gamma p}$ denotes the magnitude of the 
$\gamma$ three-momentum in the c.m. frame of the $\gamma p \to \psi p$ process:
\bea
q_{\gamma p} = \frac{(s - M^2)}{2 \sqrt{s}}.
\eea
Eqs.~(\ref{eq:discel1}), (\ref{eq:discel2}) and (\ref{eq:sigmael}) then yield the parameterization for 
the $\gamma p \to \psi p$ total cross section:
\bea
\sigma (\gamma p \to \psi p) = \left( \frac{e f_\psi}{M_\psi} \right)^2 \frac{C_{el}}{2 \sqrt{s} \, q_{\gamma p}} \left( \frac{q_{\psi p}}{q_{\gamma p}} \right) \, \left( 1 - \frac{\nu_{el}}{\nu} \right)^{b_{el}}  \left( \frac{\nu}{\nu_{el}} \right)^{a_{el}}.
\label{eq:fitel}
\eea

The discontinuity across the inelastic cut, ${\rm Disc}_{inel}$, 
is related through the optical theorem to the $\psi \, p \to c \bar c X$ inelastic cross section 
$\sigma_{\psi p}^{inel}$ as~:
\bea
{\rm Disc}_{inel} T_{\psi p}(s) = 2 \sqrt{s} \, q_{\psi p} \, \sigma_{\psi p}^{inel}.   
\label{eq:discinel2}
\eea 
Using again VMD allows to relate the inelastic cross section 
$\sigma_{\psi p}^{inel}$ to the $\gamma p \to c \bar c X$ cross section, 
with an analogous relation as in Eq.~(\ref{eq:sigmael}): 
\bea
\sigma_{\psi p}^{inel} = \left( \frac{M_\psi}{e f_\psi} \right)^2  \left( \frac{q_{\gamma p}}{q_{\psi p}} \right)^2 \,  \sigma (\gamma p \to c \bar c X). 
\label{eq:sigmainel}
\eea
Eqs.~(\ref{eq:discinel1}), (\ref{eq:discinel2}) and (\ref{eq:sigmainel}) then yield the parameterization for 
the $\gamma p \to c \bar c X$ total cross section:
\bea
\sigma (\gamma p \to c \bar c X) = \left( \frac{e f_\psi}{M_\psi} \right)^2 \frac{C_{inel}}{2 \sqrt{s} \, q_{\gamma p}}  \left( \frac{q_{\psi p}}{q_{\gamma p}} \right) \, \left( 1 - \frac{\nu_{inel}}{\nu} \right)^{b_{inel}}  \left( \frac{\nu}{\nu_{inel}} \right)^{a_{inel}}.
\label{eq:fitinel}
\eea

Having fixed the imaginary part of the $\psi$-p forward scattering amplitude, we then calculate its real part 
using the subtracted dispersion relation of Eq.~(\ref{eq:disp}). 
With the knowledge of the real and imaginary parts of the 
forward scattering amplitude $T_{\psi p}$, we can determine 
the forward ($ t = 0$) differential cross section for the $\gamma p \to \psi p$ process using VMD:
\bea
\frac{d \sigma}{dt} \biggr|_{t = 0} (\gamma p \to \psi p) 
&=& \left( \frac{e f_\psi}{M_\psi} \right)^2  \left( \frac{q_{\psi p}}{q_{\gamma p}} \right)^2 \,   \frac{d \sigma}{dt} \biggr|_{t = 0} (\psi p \to \psi p) \, \nonumber \\
&=& \left( \frac{e f_\psi}{M_\psi} \right)^2  \frac{1}{64 \, \pi \, s \, q_{\gamma p}^2} \, \big| T_{\psi p}(\nu) \big|^2.
\label{eq:dsigmadt0_gapjpsip}
\eea
Note that on the {\it lhs} of Eq.~(\ref{eq:dsigmadt0_gapjpsip}) the $\gamma p \to \psi p$ 
differential cross section is obtained at the 
unphysical point $t=0$. Its experimental determination thus requires an extrapolation from $t = t_{min}$ to $t = 0$.

Our formalism has 7 parameters: 3 parameters describing the elastic discontinuity, 3 describing the inelastic discontinuity, and the 
subtraction constant $T_{\psi p}(0)$. 
We obtain the values for these 7 parameters ($T_{\psi p}(0)$, $a_{el/inel}$, $b_{el/inel}$, $c_{el/inel}$) by simultaneously fitting the available data for
$\sigma (\gamma p \to \psi p)$, $\sigma (\gamma p \to c \bar{c} X)$ and $d\sigma/dt|_{t = 0}\,(\gamma p \to \psi p)$. 
For this purpose we use the Levenberg-Marquardt algorithm~\cite{Levenberg,Marquardt} 
of the non-linear least-squares optimization procedure implemented in \textsc{MINPACK}~\cite{More:1980dr}. 
For our fits we estimate the error based on a covariance matrix for the parameters and a linear uncertainty propagation for each of the functions.
The covariance matrix ${\bold\Sigma}_p$ is obtained based on the experimental uncertainties of the data values as follows:
\be
{\bold\Sigma}_p \,=\, \left(\,\mathrm{\bold J}_f^\mathrm{T} \,\cdot\,{\bold\Sigma}_e^{-1}\,\cdot\, \mathrm{\bold J}_f \, \right)^{-1} ,\;\;\;\;
\mathrm{J}_f^{(i,j)} = \frac{\partial f}{\partial p_j} (W_i) \,,\;\;\;\;
\Sigma_e^{(i,j)} = \delta^{(i,j)}(\sigma_e^i)^2 ,
\ee
where $W_i$ is the W-value of the $i$-th data point  (with $s$ = W$^2$), 
$\sigma_e^i$ is the total experimental uncertainty of the $i$-th data-point, 
and $f$ is the fit function of interest which depends on the energy variable (W) and the set of 7 parameters ($\{p\}$).
The derivatives over the parameters are taken at their fitted values.

\begin{figure}[h]
\begin{center}
\includegraphics[width=0.65\textwidth]{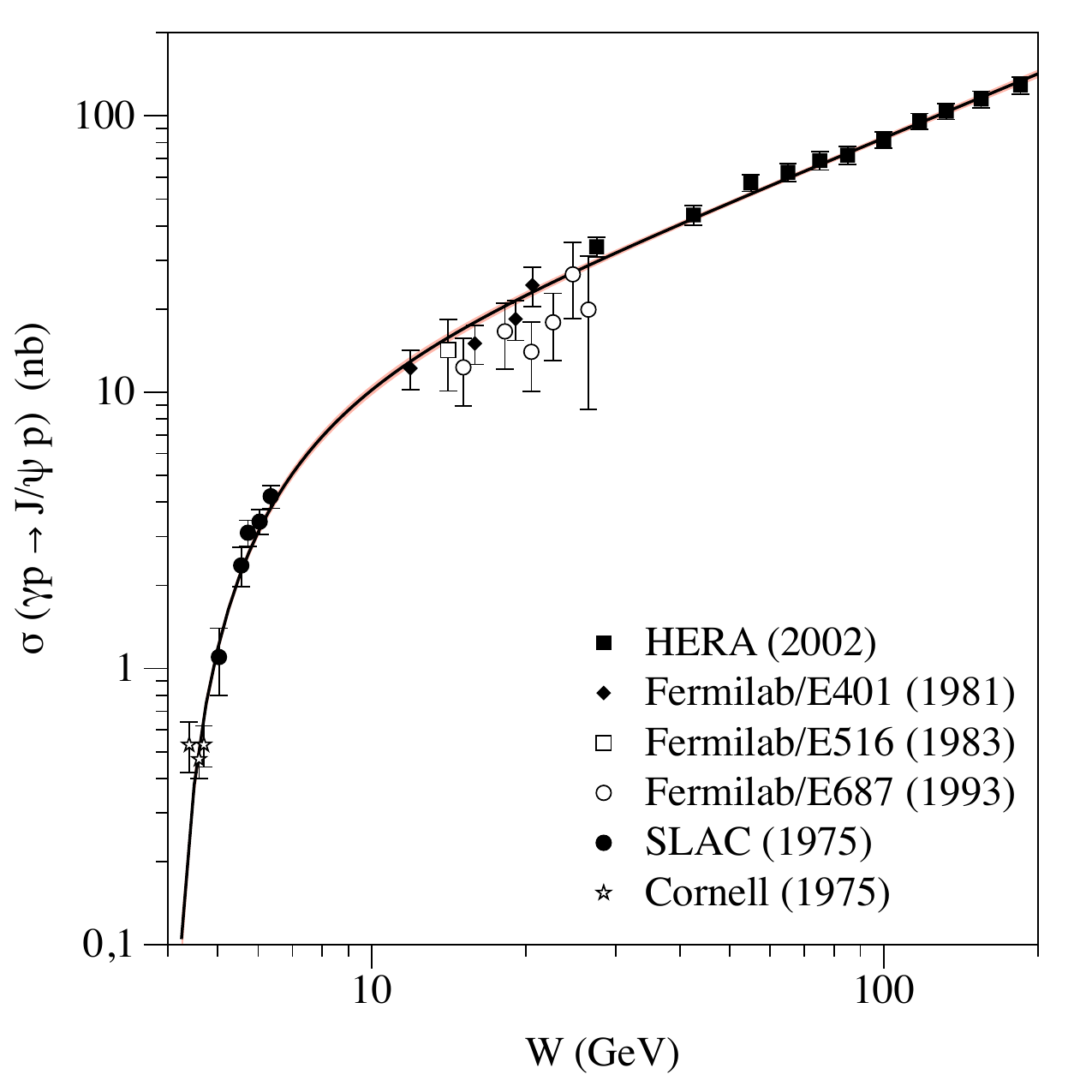}
\caption{W-dependence of the $\gamma p \to J/\psi p$ total cross section. 
The data are from Cornell~\cite{Gittelman:1975ix}, SLAC~\cite{Camerini:1975cy}, Fermilab~\cite{Binkley:1981kv,Denby:1983az,Frabetti:1993ux}, and HERA~\cite{Chekanov:2002xi}. The curve and band is the result of our global fit using Eq.~(\ref{eq:fitel}) with parameters given in Table~\ref{tab:fits} (second column, $x = el$). 
}
\label{fig:jpsi_sigmatot}
\end{center}
\end{figure}

 \begin{figure}[h]
\begin{center}
\includegraphics[width=0.65\textwidth]{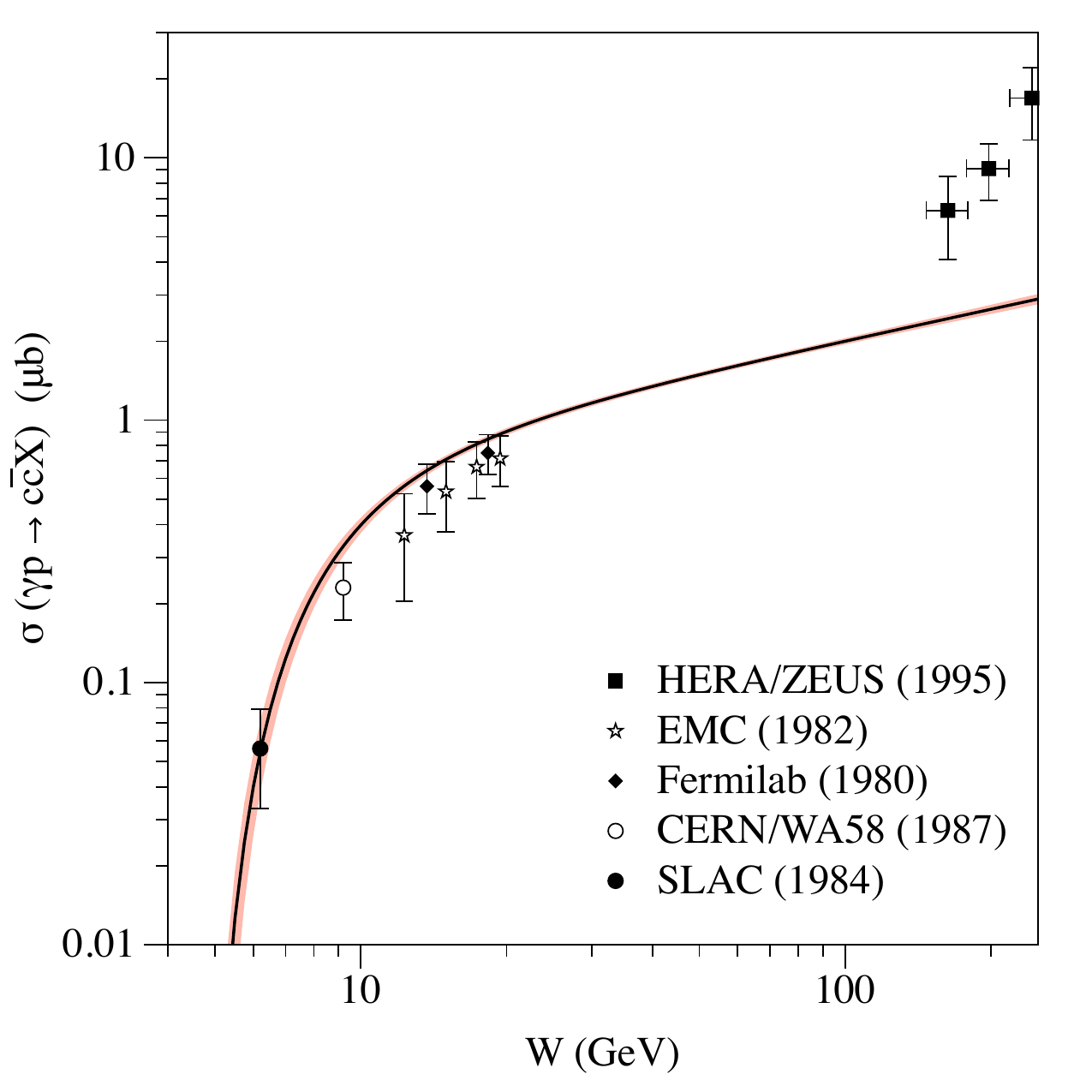}
\caption{W-dependence of the $\gamma p \to c \bar c X$ total cross section. 
The data are from 
SLAC~\cite{Abe:1983ck}, Fermilab~\cite{Clark:1980ed}, EMC~\cite{Aubert:1982tt}, 
CERN/WA58 \cite{Adamovich:1986gx}, and HERA/ZEUS~\cite{Derrick:1995sc}. 
The curve and band is the result of our global fit using Eq.~(\ref{eq:fitinel}) with parameters given in 
Table~\ref{tab:fits} (third column, $x = inel$). 
}
\label{fig:ccbar_sigmatot}
\end{center}
\end{figure}

\begin{figure}
\begin{center}
\includegraphics[width=0.65\textwidth]{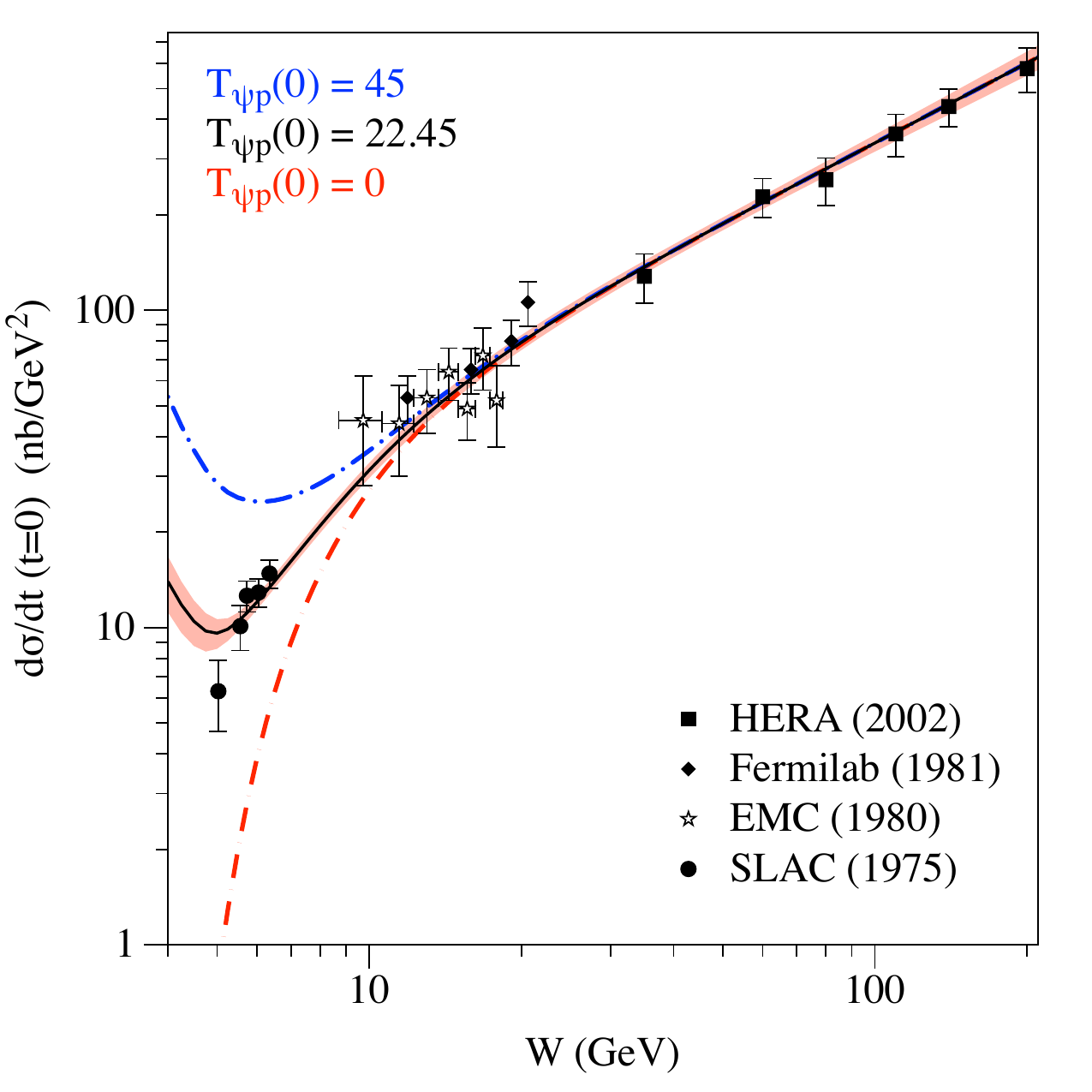}
\caption{W-dependence of the $\gamma p \to J/\psi p$ differential cross section, extrapolated to the forward direction ($t=0$), 
for different values of the subtraction constant $T_{\psi p} (0)$ in the forward $\psi$-p scattering amplitude.  
The data are from SLAC~\cite{Camerini:1975cy}, CERN/EMC~\cite{Aubert:1979ri}, 
Fermilab~\cite{Binkley:1981kv}, and HERA~\cite{Chekanov:2002xi}. 
The black curve $T_{\psi p}(0) = 22.45$ shows the best fit value, with corresponding error band resulting from our fitting procedure. 
}
\label{fig:dsigmadt0}
\end{center}
\end{figure}

We show our fit to the $\gamma p \to \psi p$ ($\gamma p \to c \bar c X$) total cross section world data  
in Fig.~\ref{fig:jpsi_sigmatot} (Fig.~\ref{fig:ccbar_sigmatot}) respectively 
as function of the c.m. energy W. 
By comparing Figs.~\ref{fig:jpsi_sigmatot} and \ref{fig:ccbar_sigmatot} 
one notices that the ratio of the inelastic over elastic cross sections 
$\sigma(\gamma p \to c \bar c X) / \sigma(\gamma \, p \to \psi \, p)$ is around a factor of 30 - 50. 
Therefore, the inelastic discontinuity dominates the determination of the forward 
$\psi$-p amplitude. Furthermore, we 
show the forward differential cross section for the $\gamma p \to \psi p$ process in 
Fig.~\ref{fig:dsigmadt0}  for three values of the subtraction constant $T_{\psi p}(0)$. 
We note that the few HERA data points for the inelastic cross section in Fig.~\ref{fig:ccbar_sigmatot} 
at the highest energies (W > 100~GeV) are not 
so well reproduced. However, these data points 
only marginally influence the fit, which in this region is mainly driven by the precise 
forward differential cross section data of Fig.~\ref{fig:dsigmadt0}.  
Our global fit yields the parameters for the elastic and inelastic discontinuities 
shown in Table~\ref{tab:fits} (second column: $x = el$, third column: $x = inel$). 
For the subtraction constant we obtain the fitted value of $T_{\psi p}(0) = 22.45 \pm 2.45$.
As an indicator of the quality of our fit, we evaluated the reduced chi-squared,
\bea
\chi^2_\mathrm{red} = \chi^2/(N_\mathrm{d}-N_\mathrm{p}),
\eea
with $N_\mathrm{d}$ being the total number of the data points we use in our fitting procedure
and $N_\mathrm{p}$ being the number of fitting parameters.
With the values of $N_\mathrm{d} = 62$ and $N_\mathrm{p} = 7$,
we get $\chi^2_\mathrm{red} = 1.36$.
Its value shows that the present fit is decent (i.e. one is not over fitting) although not perfect, 
indicating the lack of a sufficient rich database. Forthcoming more precise data especially  
in the threshold region, as expected from the planned JLab experiments~\cite{SoLID, HallB, HallC}, 
will have a strong impact on the quality of such a fit. 

We also like to note that the $\gamma p \to \psi p$ reaction in the threshold region might by modified due to the 
presence of pentaquark resonances, which have recently been reported by the LHCb Collaboration~\cite{Aaij:2015tga}. 
A new experiment at JLab~\cite{HallC} aims to search for such resonances in the $\gamma p \to \psi p$ reaction in the threshold region. If a sizable excitation strength of such pentaquark states in the $\gamma p \to \psi p$ threshold cross section would be present, it was estimated in Ref.~\cite{HallC}, based on a model calculation~\cite{Wang:2015jsa}, that this would most likely occur at larger values of $-t$, away from the forward region, and thus not influence our analysis. If, 
on the other hand, such resonances would yield sizeable excitation strength in the forward region, they could be added to the parameterization of the elastic discontinuity. The dispersion relation of Eq.~(\ref{eq:disp}) would then allow to quantify the change to the real part of the forward $\psi$-p amplitude.

\begin{table}
%\begin{center}
\begin{tabular}{c|c|c}
& $x = el$ \quad & $x = inel$ \quad  \\ 
\hline
\quad  $C_x$ \quad 
& \quad $0.10 \pm 0.01$ \quad & \quad  $20.51 \pm 1.70$ \quad \\
\hline
\quad  $b_x$ \quad 
& $1.27 \pm 0.17$ &   \quad $3.53 \pm  0.66$ \quad \\ 
\hline
\quad  $a_x$ \quad 
& $1.38  \pm 0.01$ & \quad $1.20 \pm 0.01$ \quad   \\ 
\end{tabular}
%\end{center}
\caption{Fit results for the coefficients entering the elastic discontinuity (second column, $x = el$) of Eq.~(\ref{eq:discel1}), 
and the inelastic discontinuity (third column, $x = inel$) of Eq.~(\ref{eq:discinel1}).}
\label{tab:fits}
\end{table}

\begin{figure}
\begin{center}
\includegraphics[width=0.65\textwidth]{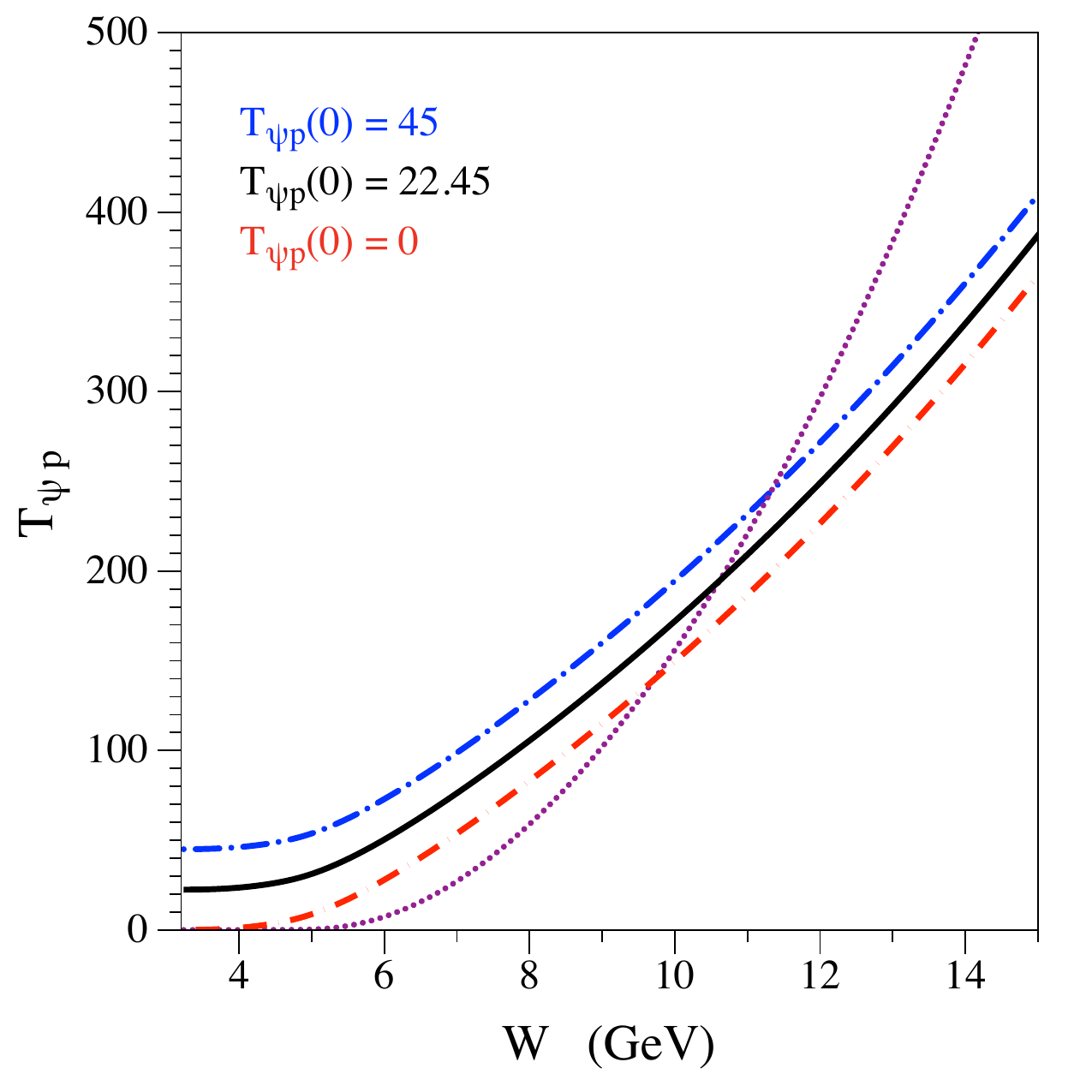}
\includegraphics[width=0.65\textwidth]{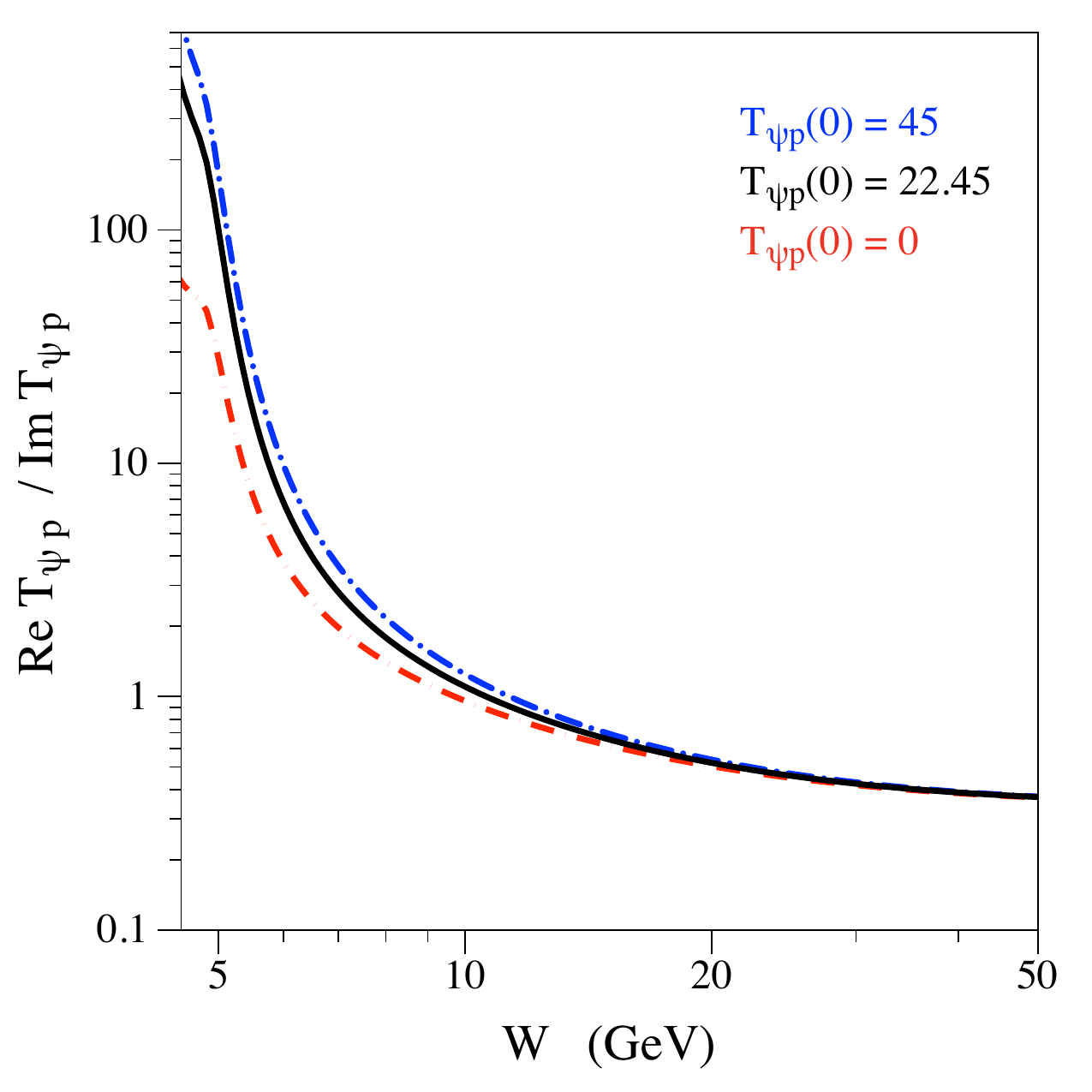}
\caption{Upper panel: Imaginary part (dotted curve) and real part of the forward scattering amplitude $T_{\psi p}$ 
as function of W.  
The real part is shown for three values of the subtraction constant as indicated on the figure.
Lower panel: corresponding ratios of real over imaginary parts.}
\label{fig:psip_psip}
\end{center}
\end{figure}

We show the real and imaginary parts of $T_{\psi p}$ in Fig.~\ref{fig:psip_psip} for our best fit value 
of the subtraction constant $T_{\psi p}(0) = 22.45$, as well as for two values around this:  $T_{\psi p}(0) = 0$ and $T_{\psi p}(0) = 45$. 
We notice that for W~$< 10$~GeV the real part dominates over the imaginary part, whereas at very high energies  (W~$\gg 10$~GeV) the amplitude is largely dominated by the imaginary part, as expected for a 
diffractive process.  

We can relate the forward $\psi$-p amplitude at threshold, 
$T_{\psi p}(\nu = \nu_{el})$, corresponding with $\sqrt{s} = M + M_\psi$, with the value of the 
$\psi$-p s-wave scattering length, $a_{\psi p}$, defined as
\bea
T_{\psi p}(\nu = \nu_{el}) = 8 \pi (M + M_\psi) \, a_{\psi p}, 
\eea
where our sign definition of $T_{\psi p}$ is fixed by Eq.~(\ref{eq:opt}). Note that in this convention, in the 
absence of a $\psi$-p bound state, a positive (negative) value of $a_{\psi p}$ 
corresponds to a positive (negative) s-wave phase shift, describing low-energy 
scattering from a weakly attractive (repulsive) potential.  
Using the dispersion relation Eq.~(\ref{eq:disp}) to relate $T_{\psi p}(0)$ with 
$T_{\psi p}(\nu_{el})$, we show the corresponding scattering lengths for three choices of the 
subtraction constant in Table~\ref{tab:scattlength}. 
Note that our best fit value $T_{\psi p}(0) = 22.45$ results in a $\psi$-p scattering length $a_{\psi p} \sim 0.05$~fm, 
which is at the lower end of the range of values estimated in the literature, ranging from 
$a_\psi = 0.05$~fm~\cite{Kaidalov:1992hd} to $a_\psi = 0.37$~fm~\cite{Sibirtsev:2005ex}.  
The value $a_{\psi p} \sim 0.05$~fm corresponds with a threshold $\psi$-p total cross section of $\sigma_{\psi p} \sim 0.3$~mb.

In a linear density approximation, the $\psi$-p scattering length $a_{\psi p}$ can be related to the $\psi$ 
binding energy in nuclear matter, $B_\psi$, corresponding with the depth of the potential well seen by $\psi$ in nuclear matter, as~\cite{Kaidalov:1992hd}
\begin{eqnarray}
B_\psi \simeq \frac{8 \pi (M + M_\psi) a_{\psi p}}{4 M M_\psi} \, \rho_{nm},
\label{eq:nmbe}
\end{eqnarray}
where $\rho_{nm} \simeq 0.17$~fm$^{-3}$ denotes the nuclear matter density. We show the $B_\psi$ values corresponding with the three values of $T_{\psi p}(0)$ considered in our calculations in Table~\ref{tab:scattlength} (last column).
Our best fit value $a_{\psi p} \sim 0.05$~fm thus corresponds to a $\psi$ binding energy in nuclear matter of 
$B_\psi \sim 3$~MeV.

\begin{table}
%\begin{center}
\begin{tabular}{c|c|c|c}
\quad $T_{\psi p}(0)$ \quad &  \quad $T_{\psi p}(\nu = \nu_{el})$  \quad & \quad $a_{\psi p}$ (in fm) \quad  & \quad $B_{\psi}$ (in MeV) \quad \\ 
\hline
\quad  $0$ \quad & $1.30$ & \quad  $0.003$ \quad & \quad  $0.2$ \quad  \\
\hline
\quad  $22.45 \pm 2.45$ \quad & $23.74 \pm 2.59$ & \quad $0.046 \pm 0.005$ \quad & \quad  $2.7 \pm 0.3$ \quad \\ 
\hline
\quad  $45$ \quad & $46.30$ & \quad $0.090$ \quad & \quad  $5.2$ \quad  \\ 
\end{tabular}
%\end{center}
\caption{Values of 
the subtraction term $T_{\psi p}(0)$ (first column), 
the corresponding values of the threshold amplitude $T_{\psi p}(\nu_{el})$ (second column), 
the corresponding $\psi$-p s-wave scattering lengths $a_{\psi p}$ (third column), 
and the corresponding $\psi$-nuclear matter binding energy $B_\psi$, according to Eq.~(\ref{eq:nmbe}) (fourth column). 
}
\label{tab:scattlength}
\end{table}

\section{Forward-backward asymmetry in the $\gamma \, p \to \psi \, p \to e^-e^+ \, p$ process}

\begin{figure}[h]
\includegraphics[width=0.4\columnwidth]{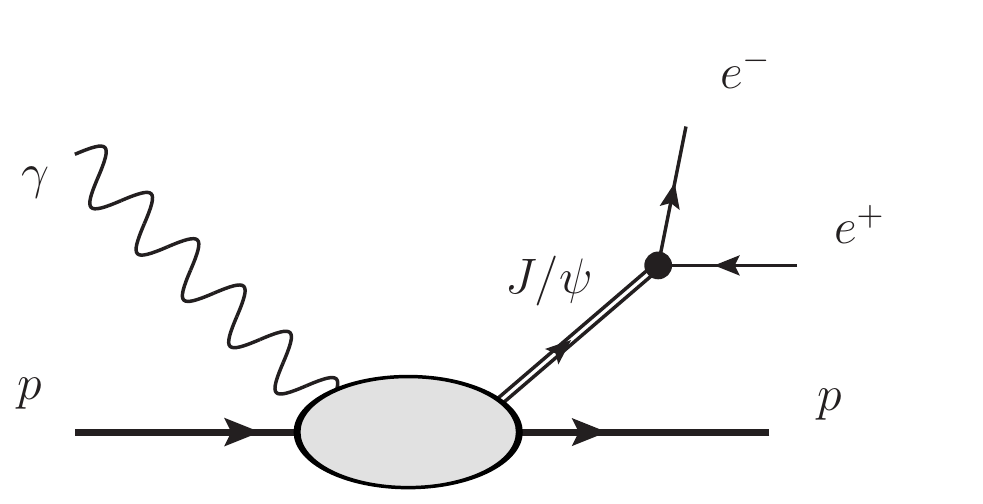}
\hspace{5mm}
\includegraphics[width=0.4\columnwidth]{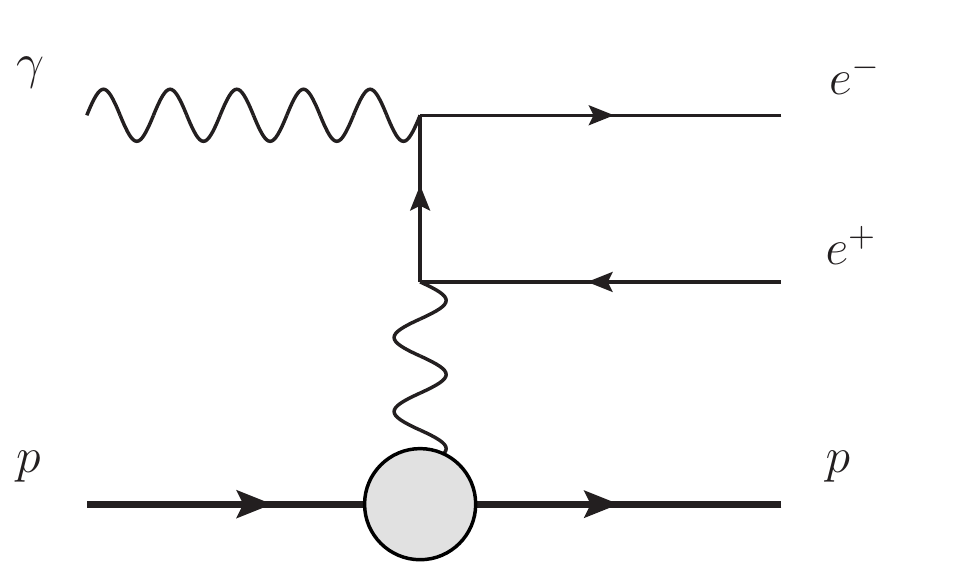}
\caption{Dilepton ($e^+e^-$) photoproduction through $J/\psi$ (left) and Bethe-Heitler (right) processes.}
\label{fig:Born_BH_diag}
\end{figure}

The value of $T_{\psi p}(0)$ extracted in the previous section is mainly sensitive to the forward differential cross section
of the $\gamma \, p \to \psi \, p$ process in the threshold region. 
The experimental access of 
the $\gamma \, p \to \psi \, p$ process proceeds through the reconstruction of the decay $\psi \to e^- e^+$ 
(or $\psi \to \mu^- \mu^+$), shown in Fig.~\ref{fig:Born_BH_diag} (left). 
In the threshold region in the forward direction (at a small value of $-t$), one may have a significant interference with the 
competing Bethe-Heitler mechanism, shown in Fig.~\ref{fig:Born_BH_diag} (right), which results in the same final state. 
We will study this interference in this section, and exploit it to find an observable which depends linearly on $T_{\psi p}(0)$.  

For this purpose, we  will calculate the observables for the process 
$\gamma ( q, \lambda) + p(p, s_p) \to e^-(l_-, s_-)  + e^+(l_+, s_+) + p(p^\prime, s^\prime_p)$ (Fig.~\ref{fig:Born_BH_diag}, left), 
with $q$, $p$, $p^\prime$, $l_-$, $l_+$ the four-momenta of initial photon, initial proton, final proton, final electron and positron respectively; and where $\lambda$, $s_p$, $s^\prime_p$, $s_-$, $s_+$ are the corresponding helicities.   
In the following expressions 
we denote the average nucleon four-momentum by $P = (p^\prime + p)/2$, 
the four-momentum of the $e^- e^+$ pair as $q^\prime = l_- + l_+$, 
the invariant mass squared of the di-lepton pair as $M^2_{ll} = q^{\prime \, 2}$, 
and indicate the squared 
momentum transfer between initial and final protons as $t = (p^\prime - p)^2$. 

For small values of $-t$, the near-forward  invariant
amplitude for the $\gamma \, p \to \psi \, p \to e^-e^+ \, p$ process is given by
\bea
{\cal M}_{\psi} &\simeq& \frac{i e^3}{q^{\prime \, 2}} 
\, \frac{f_\psi^2}{2 M} \frac{1}{q^{\prime \, 2} - M_\psi^2 + i M_\psi \Gamma_\psi} \, T_{\psi p}\left(\nu = \frac{1}{2}(s - M_\psi^2 - M^2) \right) \, \nonumber \\
&\times& \varepsilon_\mu(q, \lambda) \cdot \bar u(l_-, s_-) \gamma_\nu v(l_+, s_+)  \, 
\nonumber \\
&\times&  \bar N(p^\prime, s^\prime_p) \left\{ 
\left( g^{\mu \nu} - \frac{q^{\prime \mu} q^\nu}{q \cdot q^\prime}  \right) 
+ \frac{q \cdot q^\prime}{(q \cdot P)^2} 
\left( P^\mu - \frac{q \cdot P}{q \cdot q^\prime} q^{\prime \, \mu}\right) 
\left( P^\nu - \frac{q \cdot P}{q \cdot q^\prime} q^{\nu}\right) 
\right\} N(p, s_p),
\label{eq:jpsi}
\eea
where $T_{\psi p}(\nu)$ is the forward $\psi p$ elastic scattering amplitude discussed above, 
and $\Gamma_\psi = 92.9 \pm 2.8$~keV is the total $\psi$ width. 
Furthermore in Eq.~(\ref{eq:jpsi}), $u (v)$ denote the $e^- (e^+)$ spinors, $N$ denotes the nucleon spinors, and 
$\varepsilon_\mu$ is the initial photon polarization vector.  
The expression of Eq.~(\ref{eq:jpsi}) corresponds to a near-forward approximation, as it involves the $\psi$-p amplitude 
$T_{\psi p}$ at $t = 0$, where terms of order $-t / s$ are neglected. 

The $\gamma p \to \psi p \to e^- e^+ p$ cross section, differential in $t$, $M_{ll}^2$, and the electron solid angle $d \Omega^{e^- e^+ cm}$ in the c.m. frame of the di-lepton pair is given by~:
\begin{eqnarray}
\frac{d \sigma}{dt dM_{ll}^2 d \Omega^{e^- e^+ cm}} = \frac{1}{(2 \pi)^4} \frac{1}{64 (s - M^2)^2} \cdot \frac{1}{4} 
\sum_{\lambda} \sum_{s_p} \sum_{s^\prime_p} \sum_{s_-} \sum_{s_+}  \big| {\cal M}_{\psi} \big|^2,
\label{eq:diffcrossee} 
\end{eqnarray}
with amplitude ${\cal M}_{\psi}$ given by Eq.~(\ref{eq:jpsi}). 
We performed the check that when integrating Eq.~(\ref{eq:diffcrossee}) 
over the electron solid angle and di-lepton invariant mass, one obtains:
\begin{eqnarray}
\int dM_{ll}^2  \int d \Omega^{e^- e^+ cm}    \frac{d \sigma}{dt dM_{ll}^2 d \Omega^{e^- e^+ cm}}  \biggr|_{t = 0}  &=& 
\left( \frac{\Gamma_{\psi \to ee}}{\Gamma_\psi} \right)  \cdot \frac{d \sigma}{dt} \biggr|_{t = 0} (\gamma p \to \psi p), 
 \end{eqnarray}
with $d \sigma / dt |_{t = 0}$ being the $\gamma p \to \psi p$ forward differential cross section given by Eq.~(\ref{eq:dsigmadt0_gapjpsip}), 
multiplied by the $\Gamma_{\psi \to ee}/\Gamma_\psi$ branching ratio.

An irreducible background to the above $\gamma p \to \psi p \to e^- e^+ p$ process arises from the 
Bethe-Heitler (BH) process (Fig.~\ref{fig:Born_BH_diag}, right).
The BH invariant amplitude, contributing to the  $\gamma \, p \to e^-e^+ \, p$ reaction, is given by
\bea
{\cal M}_{BH} &=& \frac{i e^3}{t} 
 \varepsilon^\mu(q, \lambda) \cdot  \bar N(p^\prime, s^\prime_p) \Gamma^\nu N(p, s_p) 
 \nonumber \\
&\times&  \bar u(l_-, s_-) \left\{ 
\gamma_\mu \frac{\gamma \cdot (l_- - q) + m}{- 2 l_- \cdot q} \gamma_\nu 
+ \gamma_\nu \frac{\gamma \cdot (q - l_+) + m}{- 2 l_+ \cdot q} \gamma_\mu 
\right\} v(l_+, s_+)  ,
\label{eq:bh}
\eea
with nucleon vertex given by
\bea
\Gamma^\nu = F_1(t) \gamma^\nu + F_2(t)  \frac{i \sigma^{\nu \alpha} (p^\prime - p)_\alpha}{2 M},
\eea
where $F_1 (F_2)$ are the Dirac (Pauli) proton electromagnetic form factors, 
which we take from the recent fit of elastic e-p scattering data of Refs.~\cite{Bernauer:2010wm, Bernauer:2013tpr}.   
In the presence of the Bethe-Heitler, the amplitudes  ${\cal M}_{\psi}$ and ${\cal M}_{BH}$ interfere, and the differential 
cross section is obtained by an analogous expression as Eq.~(\ref{eq:diffcrossee}), with the replacement 
${\cal M}_{\psi} \to {\cal M}_{\psi} + {\cal M}_{BH}$.
 
 \begin{figure}[h]
\begin{center}
\includegraphics[width=0.6\textwidth]{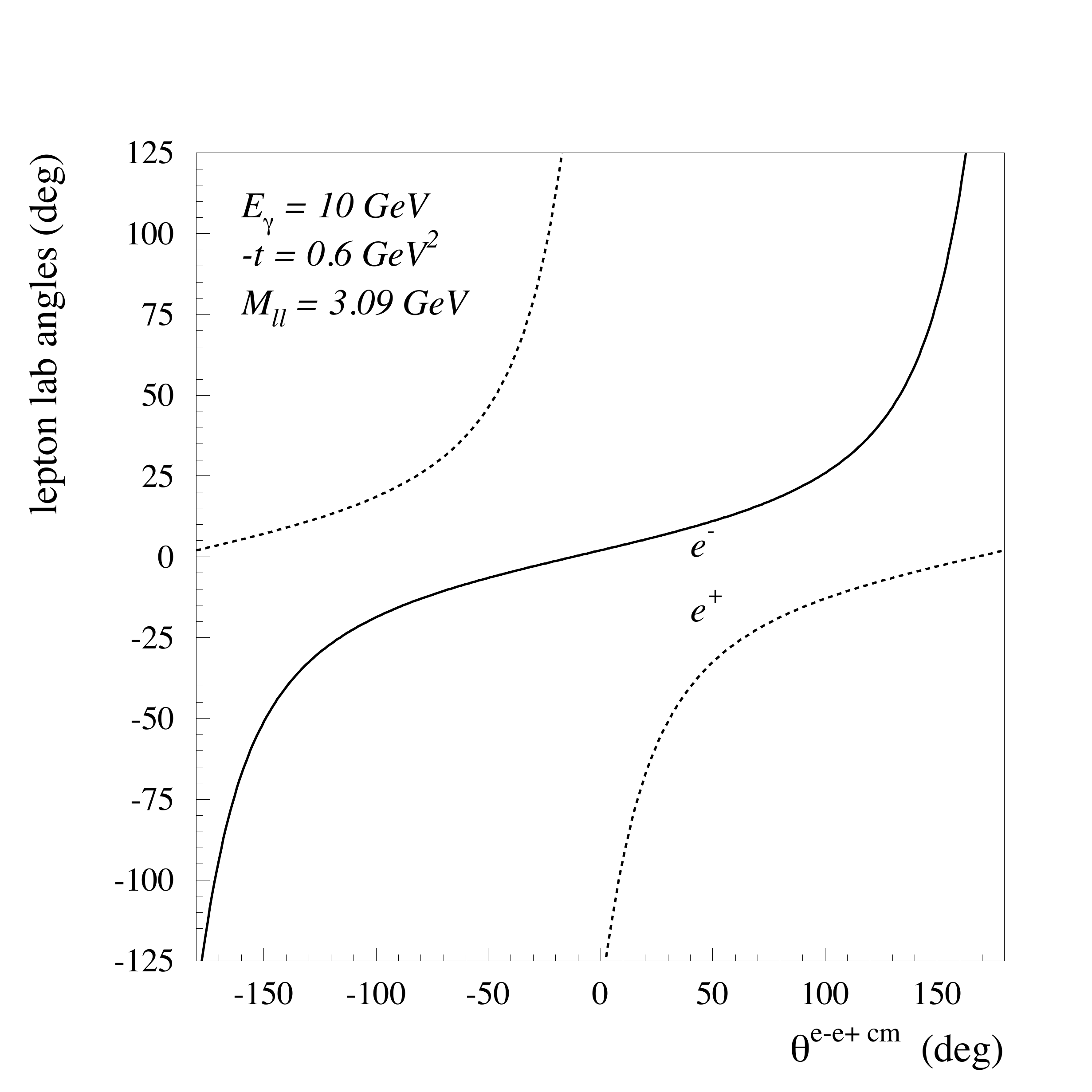}
\caption{Di-lepton angles in the {\it lab} frame as function of the electron angle 
$\theta^{e-e+ cm}$, defined in the $e^-e^+$ c.m. frame. 
Solid (dashed) curves denote the $e^- (e^+)$ {\it lab} angles. }
\label{fig:fvec_ega10}
\end{center}
\end{figure}
 
We will next study this interference between $\psi$-production and Bethe-Heitler mechanisms 
for typical kinematics of the $\gamma p \to e^- e^+ p$ process around the $\psi$ production threshold,  
accessible at JLab~\cite{SoLID, HallB, HallC}.   
In Fig.~\ref{fig:fvec_ega10}, we show the in-plane di-lepton 
angles in the laboratory frame as function of the electron angle 
$\theta^{e^- e^+ cm}$, defined in the $e^-e^+$  c.m. frame, 
around the $\psi$ resonance for JLab kinematics.

\begin{figure}
\begin{center}
\includegraphics[width=0.65\textwidth]{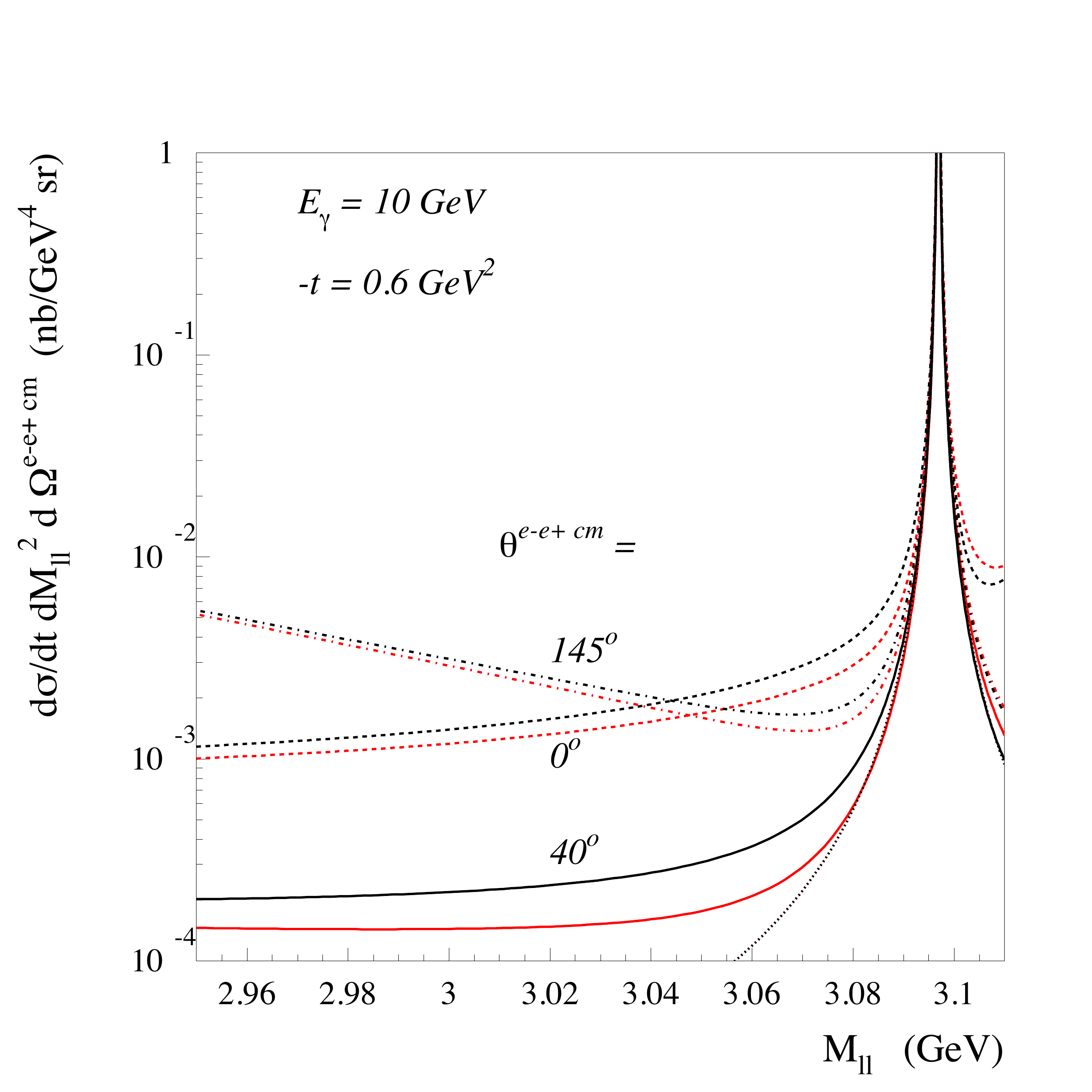}
\caption{Cross section for the in-plane 
$\gamma \, p \to  e^-e^+ \, p$ process differential in $t$, $M_{ll}^2$, and electron solid angle 
$\Omega^{e^- e^+ cm}$ in the di-lepton c.m. frame, as function of the di-lepton mass $M_{ll}$ around the 
$\psi$ resonance, and for different values of the electron angle in the di-lepton c.m. frame. 
The dotted curve denotes the $\psi$ contribution. The other curves are the Bethe-Heitler + $\psi$ results according to Eqs.~(\ref{eq:jpsi},\ref{eq:bh}) for $T_{\psi p}(0) = 22.45$. Lower red (upper black) curves show the forward (backward) angle cross sections.}
\label{fig:tcs_ega10_mll}
\end{center}
\end{figure}

Fig.~\ref{fig:tcs_ega10_mll} shows the differential cross section for the 
$\gamma p \to e^- e^+ p$ process around the $\psi$ resonance. 
For selected values of the lepton angle ($\theta^{e-e+ cm}$), we show the $\psi$ + BH cross sections 
(denoted by forward angle cross sections), and compare them with the $\psi$ + BH cross sections for the 
corresponding opposite lepton angle value ($\theta^{e-e+ cm} - 180^\circ$, denoted by backward angle cross sections).  One notices a sizable interference in the lower tail of the $\psi$ resonance.

We can directly access the real part of the $\psi$ amplitude by exploiting this interference with the BH amplitude 
through the forward-backward asymmetry $A_{FB}$ in the c.m. system of the di-lepton pair. 
We will consider the di-lepton pair in the scattering plane defined by the 
2-body $\gamma p \to J/\psi p$ process, and define the asymmetry $A_{FB}$ as~:
\bea
A_{FB} \equiv \frac{d \sigma (\theta^{e^- e^+ cm} ) - d \sigma (\theta^{e^- e^+ cm} - 180^\circ)}{d \sigma (\theta^{e^- e^+ cm} ) + d \sigma (\theta^{e^- e^+ cm} - 180^\circ)}, 
\eea 
where $d \sigma$ stands for $d \sigma / dt dM_{ll}^2 d \Omega^{e^- e^+ cm}$. 
This observable is zero for both the BH process and the $J/\psi$ process separately. 
Its non-zero value accesses the real part of the product of BH and $J/\psi$ amplitudes.

\begin{figure}
\begin{center}
\includegraphics[width=0.65\textwidth]{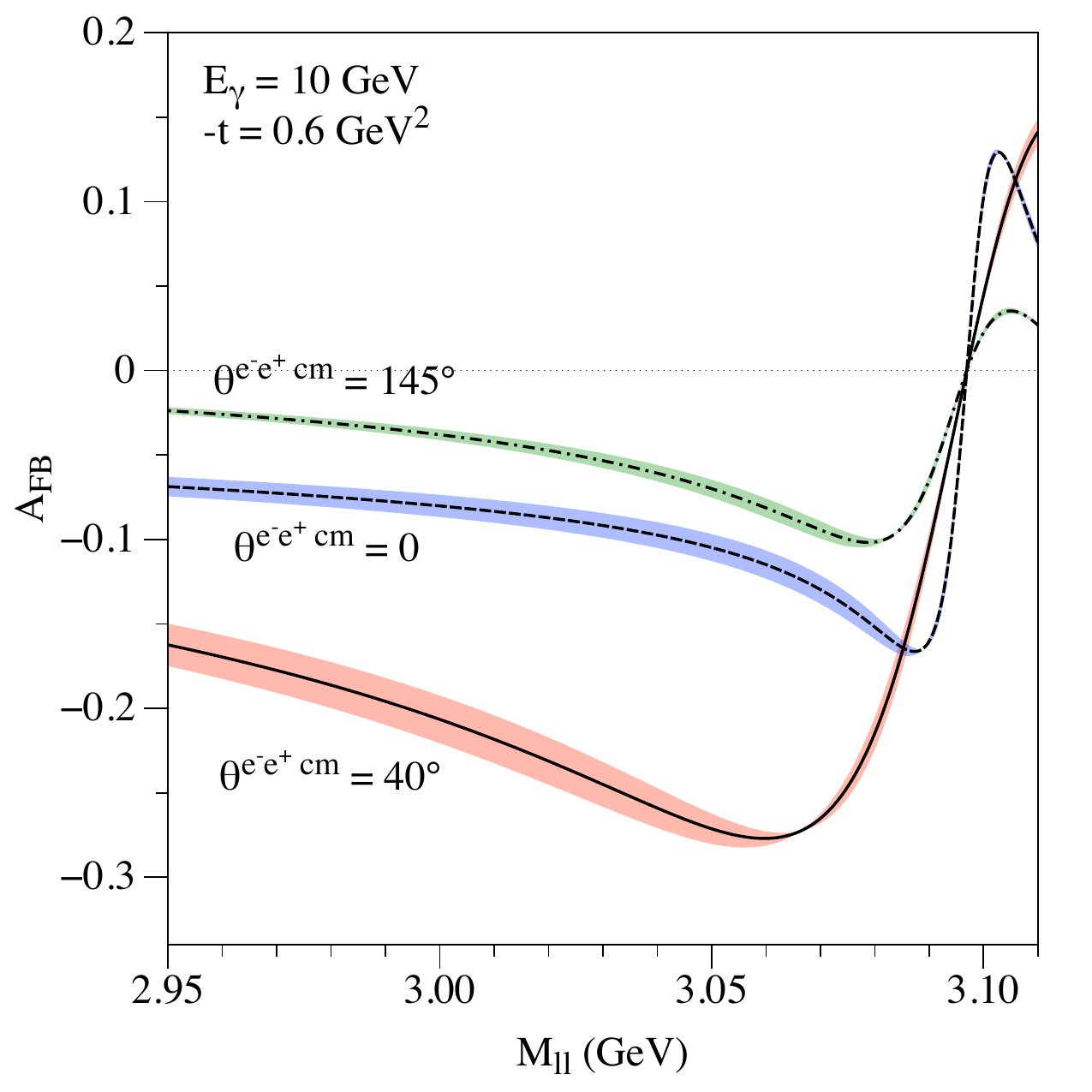}
\includegraphics[width=0.65\textwidth]{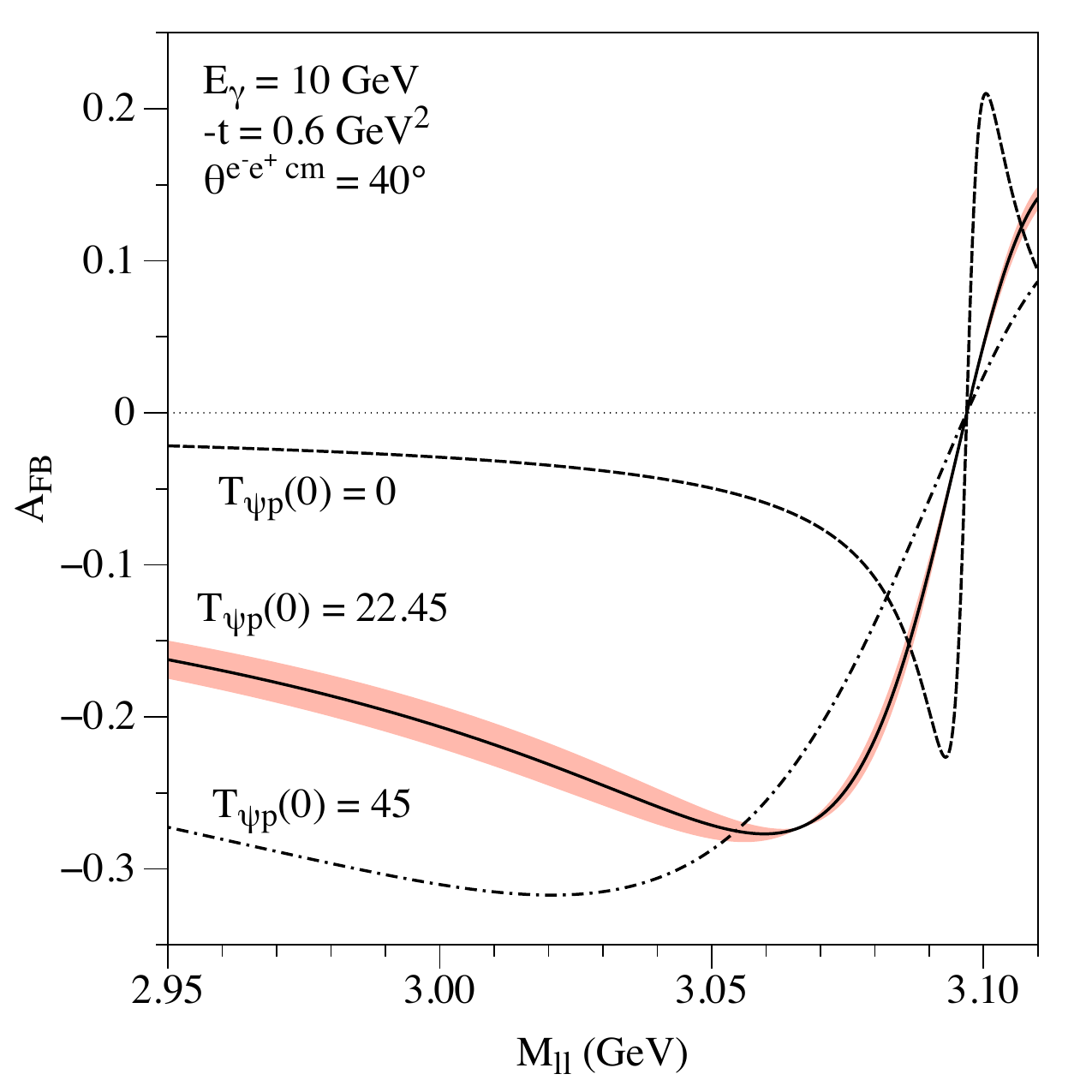}
\caption{Forward-backward asymmetry for the $\gamma \, p \to e^- e^+ \, p$ process as function of the di-lepton mass $M_{ll}$ around the $\psi$ resonance. 
Dotted curve (corresponding with $A_{FB} = 0$): $\psi$ contribution only. 
The other curves are the Bethe-Heitler + $\psi$ results. 
Upper panel: Results for $T_{\psi p}(0) = 22.45$ and for different values of the electron angle in the di-lepton c.m. frame. 
Lower panel: Results for $\theta^{e^- e^+ cm} = 40^\circ$ and for different values of $T_{\psi p}(0)$. 
The bands represent the error resulting from our cross section fitting procedure.}
\label{fig:tcs_ega10_fb}
\end{center}
\end{figure}

Fig.~\ref{fig:tcs_ega10_fb} shows the forward-backward asymmetry 
$A_{FB}$ and its sensitivity on the subtraction constant $T_{\psi p}(0)$ in the kinematics of the JLab experiments. 
We notice that the asymmetry $A_{FB}$ can reach values around $-25$~\% in these kinematics for the value 
$T_{\psi p}(0) = 22.45$ which was obtained in the fit above. In Fig.~\ref{fig:tcs_ega10_fb}, 
we apply the same procedure of linear error propagation, mentioned above for the case of the cross-section fits, 
to the asymmetry. Following this approach, the error of the asymmetry is estimated based on the first-order derivatives over the parameters:
\be
\sigma_{A_{FB}}^2 =
\frac{\partial A_{FB}}{\partial\bold p}
\cdot {\bold\Sigma}_p
\cdot \frac{\partial A_{FB}}{\partial\bold p} ,
\ee
where derivatives are taken at the fitted values of the parameters. 
We remark that in the particular case of the asymmetry all the derivatives become zero at particular values of kinematical variables ---
the values at which the Bethe-Heitler cross-section becomes equal to the $\psi$ cross-section. This is why our error bands show nodes at specific values of $M_{ll}$. 
One notices from Fig.~\ref{fig:tcs_ega10_fb} (lower panel) that away from the $\psi$ resonance position, 
the asymmetry $A_{FB}$ depends in good approximation 
linearly on the value of $T_{\psi p}(0)$. It therefore provides a very sensitive observable to extract 
$T_{\psi p}(0)$. A future measurement of $A_{FB}$ at Jefferson Lab~\cite{SoLID, HallB, HallC} 
may thus provide us with a clean way to extract $T_{\psi p}(0)$, or equivalently $a_{\psi p}$, and provide a 
cross-check of the value obtained from the cross section fits, as discussed in Section~2.

\section{Conclusions}

In this work, we provided an updated phenomenological analysis of the forward $\psi$-p scattering amplitude within a dispersive framework. Using VMD, we related the imaginary part of the forward $\psi$-p scattering amplitude to 
$\gamma p \to \psi p$ and $\gamma p \to c \bar c X$ cross section data.  
Furthermore, we calculated its real part through a once-subtracted dispersion relation. 
In our framework, the 6 parameters describing the discontinuities, and the one subtraction constant are obtained from a global fit to both total and forward differential hidden and open charm photo-production cross sections.  
This fit allowed us to extract a value for the 
spin-averaged s-wave $\psi$-p scattering length $a_{\psi p} = 0.046 \pm 0.005$~fm, which can be translated into a 
$\psi$ binding energy in nuclear matter of $B_\psi = 2.7 \pm 0.3$~MeV. 

Starting from this $\gamma p \to \psi p$ amplitude we then calculated the $\gamma p \to \psi p \to e^- e^+ p$ process.  This allowed us to estimate the forward-backward asymmetry for the $\gamma p \to e^- e^+ p$ process around the $\psi$ resonance,  which results from interchanging the leptons in the interference between the 
$\psi$ production mechanism and the competing Bethe-Heitler mechanism. 

This forward-backward asymmetry, which accesses the real part of the $\psi$-p amplitude, displays to good approximation a linear dependence on $a_{\psi p}$, away from the $\psi$ resonance position. 
Using the forward $\gamma p \to \psi p$ 
amplitude obtained from our fitting procedure,  we estimated that this forward-backward asymmetry can 
reach values around $-25$~\% for forthcoming $\psi$ threshold electro- and photo-production experiments at Jefferson Lab.  
Such forthcoming measurements can thus lead to a refined extraction of the $\psi$-p scattering length $a_{\psi p}$, 
and better constrain the $\psi$ binding energy in nuclear matter.

\section*{Acknowledgements}
We thank Zein-Eddine Meziani for very useful discussions. 
This work was supported by the Deutsche Forschungsgemeinschaft (DFG) 
in part through the Collaborative Research Center [The Low-Energy Frontier of the Standard Model (SFB 1044)], and in part through the Cluster of Excellence [Precision Physics, Fundamental
Interactions and Structure of Matter (PRISMA)].

\end{document}